\newlength{\extralineskip}
\documentclass[12pt]{article}

\usepackage{latexsym}
\usepackage{graphicx}
\usepackage{caption2}
\usepackage{psfrag}
\usepackage{amsmath}
\input epsf

\begin{document}
\begin{titlepage}
\begin{flushright}
          \begin{minipage}[t]{12em}
          \large UAB--FT--581\\
                 December 2005
          \end{minipage}
\end{flushright}
\vspace{\fill}

\vspace{\fill}

\begin{center}
\baselineskip=2.5em

{\large \bf Scalar Dark Matter and Cold Stars}
\end{center}

\vspace{\fill}

\begin{center}
{\bf  J. A. Grifols }\\
\vspace{0.4cm}
     {\em Grup de F\'\i sica Te\`orica and Institut de F\'\i sica
     d'Altes Energies\\
     Universitat Aut\`onoma de Barcelona\\
     08193 Bellaterra, Barcelona, Spain}
\end{center}
\vspace{\fill}

\begin{center}
\large Abstract
\end{center}
\begin{center}
\begin{minipage}[t]{36em}
In a medium composed of scalar particles with non-zero mass, the
range of Van-der-Waals-type scalar mediated interactions among
nucleons becomes infinite when the medium makes a transition to a
Bose-Einstein condensed phase. We explore this phenomenon in an
astrophysical context. Namely, we study the effect of a scalar
dark matter background on the equilibrium of degenerate stars. In
particular we focus on white dwarfs and the changes induced in
their masses and in their radii.

\end{minipage}
\end{center}

\vspace{\fill}

\end{titlepage}

\clearpage

\addtolength{\baselineskip}{\extralineskip}

Dark matter is one of the outstanding problems in modern cosmology
\cite {raff}. Evidence for it comes from a multiplicity of sources
and distance scales \cite {DM,DM2}. The nature and composition of
dark matter might be diverse. It might well be that each different
distance scale has its predominant type of dark matter. Several
candidates for dark matter have been postulated \cite {oli}. The
list includes the lightest supersymmetric particle (e.g., the
neutralino) and the axion. They might be heavy, like the
neutralino or light, like the axion. They might constitute the
bulk of dark matter, like WIMPS (weakly interacting massive
particles) in CDM (cold dark matter) scenarios or be just a small
fraction of it, like neutrinos (they have mass and thus they
constitute a fraction of the mass in the universe). Because scalar
particles, such as the axion or the Higgs boson, are fundamental
ingredients of the standard model (SM) of elementary particle
physics and completions thereof, scalars might constitute also a
natural ingredient of the dark stuff in the universe.

 The characteristic feature of dark matter in general is precisely the
fact that it only interacts weakly with ordinary matter and
therefore manifests itself mainly gravitationally. So far,
evidence of dark matter is indirect. Nevertheless, direct searches
that rely on the (feeble) interaction of dark particles such as
WIMPS with ordinary matter are under way and there is hope that
sufficiently sensitive experiments will eventually lead to
positive results \cite {wimp}. But there might be other effects of
the feeble interactions of dark matter with ordinary matter that
show up in other settings. In this paper we explore the
consequences of scalar dark matter in the astrophysical
environment of compact stellar objects.

Suppose very generally that in a grander unified scheme beyond the
SM, new scalar particles carrying a new conserved quantum number
$Q$ do exist. If, at some high energy scale, those putative
scalars share fundamental interactions with ordinary quarks and
with new fields associated to the high energy scale, then, at low
energies, the 2-body t-channel-exchange of these scalars will
generate feeble residual spin independent forces (i.e. Van der
Waals type dispersion forces)\cite {fein,ferrer1} among nucleons
that add coherently over unpolarized bulk matter and extend over
distances on the order of the Compton wavelength of the mediating
scalars. Furthermore, if matter is embedded in a Bose-Einstein
condensate composed of these scalars, then the dispersion force
becomes infinitely ranged, i.e. it extends over far larger
distances than the Compton wavelength. This phenomenon was first
discovered in \cite {ferrer2} and exploited in \cite {ferrer3} in
a cosmological setting.

Our starting point is the assumption that galactic dark matter is
composed at least in part by (light) scalar particles. We focus
our attention in a region where ordinary matter has accreted and
evolved into a white dwarf and the condensed scalars have been
gravitationally trapped by this stellar matter. So, we start with
an ideal boson gas in a gravitational field. The gas is at a
temperature $T\gg m$ where $m$ is the mass of the scalar. The gas
is relativistic. The statistical mechanics of a free ideal boson
gas has been given in \cite {haber}. Here, we shall describe the
statistical mechanics of Bose-Einstein condensation of a
relativistic boson gas in an external field $U(x,y,z)$. For
condensation to occur, a non zero chemical potential $\mu$ for the
gas is mandatory. We assumed before that our scalars carry a
conserved quantum number $Q$. Particles and antiparticles, i.e.
both $Q=+1$ and $Q=-1$ scalars, constitute the gas. The net total
charge $Q$ is then \footnote {For definiteness we take $Q>0$, i.e.
particles outnumber antiparticles.}
\begin {equation}
Q=\sum_{\vec{p}}\left[{1\over \exp\beta(E-\mu)-1}-{1\over
\exp\beta(E+\mu)-1}\right]
\end {equation}
with $E=\sqrt{p^2+m^2}+U(x,y,z)$ where $U(x,y,z)\geq 0$ and
vanishes at the origin, and $\beta\equiv T^{-1}$. In the continuum
limit $Q$ reads
\begin {equation}
Q=\int dV \int d^{3}\vec{p}\left[{1\over
\exp\beta(E-\mu)-1}-{1\over \exp\beta(E+\mu)-1}\right].
\end {equation}
In a central potential $U(r)$, the case we shall entertain to keep
things simple, this can be cast in the following form,
\begin {equation}
Q={1\over \pi}\int \varepsilon (E) dE \left[{1\over
\exp\beta(E-\mu)-1}-{1\over \exp\beta(E+\mu)-1}\right]
\end {equation}
with the density of states $\varepsilon (E)\equiv \int
^{\tilde{U}} r^{2}dr (E-U(r))\sqrt{(E-U(r))^{2}-m^{2}}$ and
$\tilde{U}(E)$ is such that $U(\tilde{U})=E$.

Above a critical temperature $T_{c}$ defined implicitly by the
equation
\begin {equation}
Q=Q(\mu=m,T=T_{c})
\end {equation}
the charge is correctly given by equation (2). But for $T$ below
$T_{c}$, equation (2) gives only the charge distributed in excited
states. For $T\leq T_{c}$, a macroscopic fraction of the total
charge sits in the ground state and is not included in the above
integral. The boson gas is condensed, and the particles in the
zero momentum mode constitute the condensate. For an harmonic
oscillator potential (i.e. $U(r)=\alpha r^{2}$), as would be the
case of a scalar particle inside a uniform matter distribution,
the condensation temperature can be easily calculated to be
\begin {equation}
T_{c}=\left({{8\sqrt{\pi}Q\alpha^{\frac{3}{2}}}\over
{29\zeta(\frac{7}{2})m}}\right)^{\frac{2}{7}}
\end {equation}
up to corrections ${\cal O}(m/T)$.

The condensed charge in this case is given by
\begin {equation}
Q_{0}=Q\left[1-\left(\frac{T}{T_c}\right)^{\frac{7}{2}}\right].
\end {equation}
Although Bose-Einstein condensation is condensation in
momentum-space, in the presence of a gravitational field there
will be condensation in ordinary space as well, and two separate
phases will emerge just as in ordinary gas-liquid condensation.
This fact causes the charge $Q_0$ to accumulate at $r=0$. The
thermal particles, i.e. those distributed in excited states,
bearing a kinetic energy on average large compared to their rest
mass will eventually escape from the star. Only the condensate
will be trapped at the center. Because all the charge $Q_0$ sits
at $r=0$, the density of charge grows to infinity. This is however
not realistic since the uncertainty principle prevents both
momentum and position of the condensed particles to be
simultaneously sharp. In the gravitational field of the star, the
condensate will occupy a finite region centered around $r=0$ whose
size depends on the depth and width of the potential well, and on
the mass $m$ of the scalar particles. The region will be a
macroscopic fraction of the star's volume for certain ranges of
these parameters (see below). Thus, matter in this region will be
subject to the pull of macroscopic 2-scalar exchange forces on top
of the pull of gravity. Let us give a succinct briefing on
dispersion forces in a thermal bath of the mediating scalars. The
derivation closely follows \cite {ferrer2}. Consider two nucleons
a distance $r$ apart and at rest relative to the scalar heat
reservoir. We describe the effective interaction of nucleons with
scalars with the lagrangean
\begin {equation}
{\cal L}=\frac{g}{m_N}\overline{\psi}\psi\phi\phi^\dag,
\end {equation}
where $m_N$ is the nucleon mass (chosen to set the scale of the
coefficient of the dimension$-5$ effective low energy operator)
and $\psi$ and $\phi$ are the nucleon and scalar fields,
respectively.
 It is the simplest low energy realization of the underlying fundamental
interactions (see \cite {ferrer3} for a discussion). The force
between the two nucleons arises in this approximation from the
emission at a time by one nucleon and absorbtion by the other of
two scalar quanta. We use real time finite temperature field
theory to calculate the corresponding Feynman amplitude (see e.g.
\cite {horo} for an early application of the technique). The
Fourier transform of its non-relativistic limit is the potential
we seek.

The physical clue to the nontrivial influence of the particles in
the background on the force due to 2-scalar t-channel exchange is
that while one and two $\textit{real}$ particles cannot be
exchanged in the t-channel, the exchange of one $\textit{real}$
plus one $\textit{virtual}$ particle in the t-channel is
kinematically allowed. In the case at hand, the heat bath supplies
the real particle. This mechanism contributes a piece to the
potential that reads
\begin {equation}
{\cal V}_{T}(r)=-\frac{g^{2}}{16\pi m_{N}^{2}}\frac{1}{r}\frac{1}{
V}\int dV\int
\frac{p^{2}dp/2\pi^{2}}{\sqrt{p^{2}+m^{2}}}\frac{\sin2pr}{2pr}\left[{1\over
\exp\beta(E-\mu)-1}+{1\over \exp\beta(E+\mu)-1}\right].
\end {equation}
Below $T_c$ a macroscopic fraction of the charge carried by
particles in the reservoir piles up in the zero mode state (the
condensate) and the integral over states in equation (8) just as
in equations (2) and (3) no longer correctly describes the
physical situation. To evaluate the condensate contribution to the
potential in the degenerate case ($T\leq T_c$) we next establish
its relation to $Q$. Call $n_{\pm}(p)$ the following distribution
functions,
\begin {equation}
n_{\pm}(p)={1\over \exp\beta(E-\mu)-1}\pm{1\over
\exp\beta(E+\mu)-1}.
\end {equation}
Then, for $T\leq T_c$, it holds that
\begin {equation}
n_{+}(0)=n_{-}(0)+{2\over \exp\beta(2m+U)-1}.
\end {equation}
The condensate contribution to the potential is proportional to
$\frac{1}{V}\int dVn_{+}(0)$. Since $U\geq 0$ we see from the
previous relation, equation (10), that this quantity differs from
the charge $Q_0$ in the condensate by less than $2{(\exp2\beta
m-1)}^{-1}$. Furthermore, $\beta m\ll 1$, and therefore, as long
as the net charge $Q$ is a macroscopically large number many
orders of magnitude larger than $T/m$, the factor $\frac{1}{V
}\int dVn_{+}(0)$ coincides essentially with the condensate
contribution to the charge $Q_0$. As a consequence we can write
the potential contributed by the zero momentum modes as (the
passage from discrete to continuum momentum space, or viceversa,
entails the substitution rule $V^{-1}\leftrightarrow
(2\pi)^{-3}d^{3}p$) :
\begin {equation}
{\cal V}_{cond}(r)=-\frac{g^{2}}{16\pi
m_{N}^{2}}\frac{q_0}{m}\frac{1}{r}
\end {equation}
where ${q_0}\equiv \frac{Q_0}{V}$ and the $m$ in the denominator
comes from $({p^{2}+m^{2})^{-\frac{1}{2}}}
{(2pr)^{-1}}{\sin2pr}\rightarrow m^{-1}$.

In the specific case at hand, using equations (5) and (6), we find
\begin {equation}
{\cal V}_{cond}(r)=-\frac{29\zeta (\frac{7}{2})g^{2}}{128{\pi
^{\frac{3}{2}}{m_N^{2}}}}\frac{T_c^{\frac{7}{2}}}{\alpha
^{\frac{3}{2}}V}\left
[1-\left(\frac{T}{T_c}\right)^{\frac{7}{2}}\right]\frac{1}{r}.
\end {equation}
So for a degenerate scalar background the resulting potential
behaves as $r^{-1}$, i.e. it is a potential of infinite range. On
the other hand, when the temperature of the background is above
$T_c$, the potential is cut off by the typical Yukawa damping
factor $\exp (-2mr)$ as can be seen from performing the momentum
integral in equation (8).

Finally, in a vacuum,  the potential takes the form \cite
{ferrer2}
\begin {equation}
{\cal
V}_{vac}(r)=-\frac{g^{2}m}{64\pi^{3}m_{N}^{2}}\frac{K_{1}(2mr)}{r^{2}}
\end {equation}
with $K_{1}(2mr)$, the modified Bessel function whose asymptotic
form gives rise to the characteristic exponential damping. We will
use this result later on.

Now that we have the form of the potential associated to the
$2-$scalar force among nucleons in a scalar heat bath and in
vacuum, we turn to our problem, namely to stars whose central core
(or even their whole volume) is permeated by a fluid made of
scalar particles. To be definite and simple, we shall discuss
equilibrium configurations called polytropes, i.e. with the
equation of state of the form $P=K\rho ^{\Gamma}$. More
specifically, we shall analyze idealized white dwarfs where the
Fermi pressure of a degenerate ideal gas of ultrarelativistic
electrons supports these stars against collapse (a discussion on
ordinary white dwarfs can be found in \cite {shapi}). In the core
of actual white dwarfs the nucleons are not free but bound in
nuclei. In turn, these nuclei\footnote{Typically, carbon, oxygen,
i.e. not exceeding atomic weight about 20.} tend to form crystal
like structures, i.e. they arrange themselves on a regular
lattice. Because of this circumstance the ideal Fermi gas equation
of state is modified in real life by the electrostatic
interactions among the positive ions and the electrons. The
resulting corrections can be accounted for (see e.g. \cite
{shapi}) and are generally small. Another source of departure from
the ideal gas approximation is provided by inverse $\beta-$decay
which tends to modify the number density of the electrons and thus
lessens the Fermi pressure of the electron gas. Here we are not
concerned with this correction nor with the Coulomb correction. As
far as the putative long range forces that we contemplate, the
fact that nucleons (inside nuclei) cluster on lattice sites is
irrelevant for our purposes since the potential associated to
macroscopic bulk matter results from the coherent superposition of
the underlying long distance potential created by the individual
nucleons as, incidentally, it is the case for gravity. Therefore,
to study bulk properties, the matter density is averaged over
scales much larger than the inter-ion spacing and it is smoothed
out over a macroscopically large number of nucleons. For neutron
stars, the actual equation of state deviates from an equation of
state for an ideal gas in ways that are more dramatic and much
less under control than it is the case for white dwarfs. As the
authors in \cite {shapi} put it, while for white dwarfs,
observations of masses and radii are used as confirmation of
astrophysical models, for neutron stars observations of masses and
radii are instead used to test theories of nuclear physics. It is
in this spirit that we keep things simple and do work with white
dwarfs only.

 Our aim is to study the effect on stellar equilibrium of the
condensation of the dark background particles. The hydrostatic
equilibrium in our case is described by the two equations:
\begin {equation}
\frac{1}{r^{2}}\frac{d}{dr}\left(\frac{r^{2}}{\rho}\frac{dP}{dr}\right)=-4\pi
G \left (1+\kappa \right)\rho
\end {equation}
for $0\leq r \leq L$, and
\begin {equation}
\frac{1}{r^{2}}\frac{d}{dr}\left(\frac{r^{2}}{\rho}\frac{dP}{dr}\right)=-4\pi
G\rho
\end {equation}
for $L\leq r\leq R$.

In equation (14),
\begin {equation}
\kappa\equiv \frac{g^{2}}{16\pi m_{N}^{4}G}\frac{q_0}{m}
\end {equation}
 and
$G$ in both equations is Newton's constant. The distance $L$ sets
the size of the region where the particles of the condensate are
confined. Inside a sphere of radius $L$, the inverse square law
forces of gravity and 2-scalar exchange are operative. Outside
this sphere only the pull of gravity balances the Fermi pressure.
The boundary conditions for equation (14) are $\rho(r=0)=\rho_c$
(where $\rho_c$ is the central density of the star) and
$\rho'(r=0)=0$. The boundary conditions for equation (15) follow
from the requirement that $\rho$ and $\rho'$ in equation (15)
should match $\rho$ and $\rho'$ of equation (14) at $r=L$. The
radius $R$ of the star is implicitly defined by the condition
$\rho(R)=0$ (and hence, $P(R)=0$).

In the case under scrutiny, $\Gamma =\frac{4}{3}$ and
$K=\frac{3^{\frac{1}{3}}\pi^{\frac{2}{3}}}{4m_{N}^{\frac{4}{3}}\mu_{e}^{\frac{4}{3}}}$
($\mu_e$ is the mean molecular weight per electron) \cite {shapi}.
By changing variables to dimensionless quantities $\xi$ and
$\theta$ defined as
\begin {equation}
r=a\xi \qquad \qquad \rho=\rho_c \theta^{3}
\end {equation}

where the length parameter $a\equiv\left[\frac{K}{\pi G
\left(1+\kappa\right)\rho_c^{\frac{2}{3}}} \right]^{\frac{1}{2}}$,
we can put both equations, (14) and (15), in the form:
\begin {equation}
\frac{1}{x_{1,2}^{2}}\frac{d}{dx_{1,2}}\left(x_{1,2}^{2}\frac{d\theta}{dx_{1,2}}\right)=-
\theta^{3}
\end {equation}
where $x_1\equiv \xi$ for $0\leq \xi \leq \frac{L}{a}$ and
$x_2\equiv \xi/\sqrt{1+\kappa}$ for $\frac{L}{a}\leq \xi\leq
\frac{R}{a}$. The boundary conditions in this new form now read:
\begin {equation}
\theta(x_1=0)=1 ;\qquad
\left.\frac{d\theta}{dx_1}\right|_{x_1=0}=0
\end {equation}
for the first equation, and
\begin {equation}
\theta(x_2=L/a\sqrt{1+\kappa})=\theta(x_1=L/a) ;\qquad
\left.\frac{d\theta}{dx_2}\right|_{x_2=L/a\sqrt{1+\kappa}}=\frac{d\theta}{dx_1}\sqrt{1+\kappa}
\end {equation}
for the second equation.

 These equations are to be solved numerically and we shall do so
below. We shall obtain the radius and the mass of equilibrium
configurations as a function of the size of the condensate, i.e.
as a function of the parameter $\xi_L\equiv L/a$. Before doing so
let us recall that in the ordinary case, i.e. gravity alone at
work, one gets, in the polytropic approximation, the Chandrasekhar
limit or maximum possible mass of a white dwarf. The explicit
expressions for the Chandrasekhar mass and radius are,
respectively
\begin {equation}
M_{Ch}=4\pi\left(\frac{K}{\pi G}\right)^{\frac{3}{2}}
\xi_{1}^{2}|\theta '\left(\xi_{1}\right)|
\end {equation}
and
\begin {equation}
R_{Ch}=\left(\frac{K}{\pi
G}\right)^{\frac{1}{2}}{\rho_c^{-\frac{1}{3}}}\xi_{1}
\end {equation}
where $\xi_{1}$ satisfies $\theta (\xi_{1})=0$.

The numerical integration of the corresponding differential
equation gives in this case $\xi_{1}=6.89685$ and
$\xi_{1}^{2}|\theta '\left(\xi_{1}\right)|=2.01824$ \cite {shapi}.
When plugged in equation (21), one obtains the well known value
\begin {equation}
M_{Ch}=1.457\left(\frac{2}{\mu_e}\right)^{2}M_\odot.
\end {equation}
It will be convenient for our analysis to normalize radii and
masses relative to $R_{Ch}$ and $M_{Ch}$, respectively. Trivially,
\begin {equation}
\frac{R}{R_{Ch}}=\frac{x_{2,R}}{\xi_1}
\end {equation}
where $x_{2,R}$ verifies $\theta (x_{2,R})=0$. As to the mass, it
its easy to show starting from $M=\int_{0}^{R} 4\pi r^{2}\rho dr$
that
\begin {equation}
\frac{M}{M_{Ch}}=2.01824^{-1}\left[x_{2,R}^{2}\left|\frac{d\theta}{dx_2}\right|_{x_{2,R}}-
\frac{\kappa}{(1+\kappa)^{\frac{3}{2}}}x_{1,L}^{2}\left|\frac{d\theta}{dx_1}\right|_{x_{1,L}}\right]
\end {equation}
with $x_{1,L}\equiv\xi_L\equiv \frac{L}{a}$. Sticking to ratios
$\frac{M}{M_{Ch}}$ and $\frac{R}{R_{Ch}}$ make our results less
dependent on the approximations made and make them more reliable
when trying to extract general consequences for realistic stars.

In order to proceed we first should estimate the size of the
stellar volume that can actually be filled by the particles of the
condensate. For that purpose let us study the quantum mechanics of
a scalar particle occupying the ground state in the gravitational
potential well of the star.

First of all we give the form of the potential $\Phi$. For a
spherical symmetric mass distribution,
\begin {equation}
\Phi=-\frac{Gm(r)}{r}+G\int_{0}^{r}4\pi r\rho dr
\end {equation}
where $m(r)$ is the mass inside $r$ and $\Phi(0)=\Phi'(0)=0$. In
our case, it can be explicitly cast in terms of the solution
$\theta$ of equations (18),
\begin {equation}
\Phi=4\pi G a^{2}\rho_c(1-\theta)
\end {equation}
inside the star, i.e. for $r\leq R$, while for $r\geq R$,
\begin {equation}
\Phi=4\pi G a^{2}\rho_c+MG\left(\frac{1}{R}-\frac{1}{r}\right).
\end {equation}
Since we shall deal only with the lowest energy mode, the
wavefunction $\Psi(r)$ is purely radial with no orbital part. The
radial Schr\"{o}dinger equation can be put in an equivalent
one-dimensional form with no centrifugal barrier, if we trade
$u(r)=r\Psi(r)$ for $\Psi(r)$:
\begin {equation}
\left(-\frac{1}{2m}\frac{d^{2}}{dr^{2}}+m\Phi(r)\right)u(r)=Eu(r).
\end {equation}
Because of the finiteness of $\Psi(r)$, $u(r)$ should vanish at
$r=0$. Instead of using the actual $\Phi(r)$ (given by equations
(27) and (28)), for which we have no analytical expression, we
shall introduce in equation (29) a potential of the form
$\Phi(r)=\gamma \tanh^{2}\frac{r}{\delta}$. The parameters
$\delta$ and $\gamma$, width and depth of the well, can be fitted
to give a fair approximation to the real potential. Of course,
$\delta \sim {\cal O}\left(R\right)$ and $\gamma \sim {\cal
O}\left(\Phi(\infty)\right)$ with $\Phi(\infty)$ being the
asymptotic value of equation (28). The reason for doing so is that
the Schr\"{o}dinger equation above can be given an exact analytic
solution. Indeed, the physically meaningful ground state
wavefunction that complies with the boundary conditions at $r=0$
and at infinity turns out to be, after some mathematical
manipulations,
\begin {equation}
\Psi(r)=A\left(\frac{r}{\delta}\right)^{-1}\sinh
\frac{r}{\delta}\cosh^{-2\lambda} \frac{r}{\delta}
\end {equation}
with $A$ a normalization constant and
$\lambda=\frac{1}{4}\left[-1+\sqrt{1+8m^{2}\gamma
\delta^{2}}\right]$. The corresponding ground state energy reads:
\begin {equation}
E_0=(4m\delta^{2})^{-1}\left[-5+3\sqrt{1+8m^{2}\gamma
\delta^{2}}\right].
\end {equation}
For $8m^{2}\gamma \delta^{2}\gg 1$, equations (30) and (31)
approach closely the wavefunction and the ground state energy of
an isotropic harmonic oscillator of angular frequency
$\omega=\frac{\sqrt{2\gamma}}{\delta}$. In general, however, the
ground state lies lower in energy and the wavefunction is broader
than for the oscillator well. Thus, the position uncertainty of
the harmonic oscillator is an underestimate of the uncertainty in
position of the particle in the actual potential well. We shall
use this conservative quantity in our estimates. The precise
meaning we give to the parameter $L$ is the following:
\begin {equation}
L\equiv \sqrt{\sum_i \Delta x_i^{2}}
\end {equation}
with $\Delta x_i^{2}\equiv\langle x_i^{2}\rangle-\langle
x_i\rangle^{2}$ and $\vec{r}\equiv(x_1,x_2,x_3)$. For an harmonic
oscillator, $\Delta x_i^{2}=(2m\omega)^{-1}$. In our case,
$\omega=\frac{\sqrt{2\gamma}}{\delta}$. Hence,
\begin {equation}
L=\frac{3^{\frac{1}{2}}\delta^{\frac{1}{2}}}{2^{\frac{3}{4}}\gamma^{\frac{1}{4}}m^{\frac{1}{2}}}
\end {equation}
and,
\begin {equation}
L\simeq 7\times
10^{-4}\left(1+\kappa\right)^{\frac{1}{4}}\left(\frac{\rho_c}{10^{8}g/cm^{3}}\right)^{\frac{1}{4}}\left(\frac{1eV}{m}\right)^{\frac{1}{2}}\left(\frac{R}{R_{Ch}}\right)^{\frac{1}{2}}Km
\end {equation}

if we set roughly $\delta\sim R$ and $\gamma\sim 4\pi
Ga^{2}\rho_c$.

Clearly, appreciable $L$'s $-$relative to stellar dimensions$-$
will be attained only for sufficiently large $\kappa$ and/or
sufficiently low $m$. However, neither can we go arbitrarily low
in mass if we want to keep the momentum uncertainty of the scalars
to be much less than their rest mass, nor can we increase the
strength of the potential without eventually inducing unwanted
effects in laboratory Cavendish-type experiments. We shall address
these points later. Because $L$ depends on $R$, the boundary
conditions (20) for the matching of the hydrodynamic equilibrium
equations depend on the solution of those equations. Hence, if one
solves equations (18) as a function of the $\textit{free}$
parameter $L$ and then obtains $R(L)$ through equation (24), a
stellar configuration should have a radius $R_\star$ such that the
self-consistency condition is satisfied:
\begin {equation}
R_\star=R(L=L(R_\star))
\end {equation}
where $L(R)$ stands for equation (34).

We shall now integrate numerically equations (18). Our strategy is
the following. We choose a large value for $\kappa$. We do the
numerics with this $\kappa$ fixed and letting the parameter
$\xi_L$ vary $\textit{freely}$. The output are the quantities
$x_{2,R}$, $\left|\frac{d\theta}{dx_2}\right|_{x_{2,R}}$ and
$\left|\frac{d\theta}{dx_1}\right|_{x_{1,L}}$, to be introduced in
equations (24) and (25). We repeat this procedure for various
values of $\kappa$. To comply with the self-consistency
requirement in equation (35) we check $\textit{a posteriori}$ for
each value of $\xi_L\equiv L/a$ whether a physically reasonable
mass $m$ can be found such that equation (34) is satisfied. Our
results will be then presented in the parametric form
$\left\{\frac{R}{R_{Ch}}(\xi_L),\frac{M}{M_{Ch}}(\xi_L)\right\}$.

%%%%%%%%%%%%%%%%%%%%%%%%%%%%%%%%%%%%%%%%%%%%%%%%%%%%%%%%%%%%%%%%%%%%%%%%%
%%%%%%%%%  EXAMPLE OF FIGURE  %%%%%%%%%%%%%%%%%%%%%%%%%%%%%%%%%%%%%%%%%%%
\begin{figure}

\centering \psfrag{a}{$R/R_{Ch}$} \psfrag{b}{$M/M_{Ch}$}
\includegraphics[width=5in]{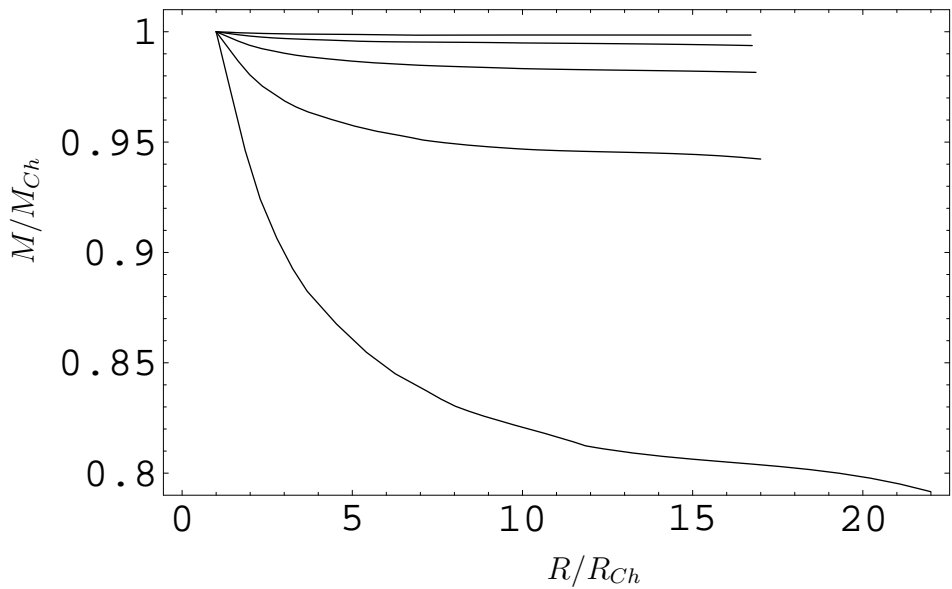}
%\hspace{1cm}
%\includegraphics[width=2.3in]{fit4.eps}
%\includegraphics[width=2.3in]{fit6.eps} \hspace{1cm}
%\includegraphics[width=2.3in]{fit8.eps}
\caption{From top to bottom: equilibrium configurations for
$1+\kappa=10^{6}$, $10^{5}$, $10^{4}$, $10^{3}$ and $10^{2}$,
respectively. Along the curves, $\xi_L$ grows from $0$ at point
$(1,1)$ to $2$ at the opposite end of the curves. }\label{ZZZ}
\end{figure}

It follows from the analysis that our stars tend to fall in two
qualitatively distinct categories. Those whose mass is somewhat
below the Chandrasekhar mass but their radius is larger than the
Chandrasekhar radius constitute a first group. In the second
category we find stars with very small masses and radii. In the
first case, the star adjusts itself such that only a small core of
the star is filled by the scalar condensate whereas in the second
case, equilibrium is attained with a large fraction of the star
occupied by the condensate at the expense of a drastic reduction
in mass and radius. The first class corresponds to low values of
the parameter $\xi_L$ (roughly up to $\xi_L\approx 2$) and the
second class corresponds to the higher segment of the $\xi_L$
range (up to $\xi_1=6.89685$ ; recall that $\xi_L$ varies from 0
to $\xi_1$). For the purpose of illustration we present the two
domains in two separate $(R,M)$-plots. In figure 1 we show our
results for $1+\kappa=10^2, 10^3, 10^4, 10^5$ and $10^6$ as a
function of the parameter $0\leq \xi_L\leq2$. In figure 2 we
display the results as a function of $3\leq \xi_L\leq\xi_1$. In
this second plot we draw a single curve. This is because it turns
out that, for large $\kappa$ and $\xi_L\geq3$, the radius and the
mass scale as $(1+\kappa)^{-\frac{1}{2}}$ and
$(1+\kappa)^{-\frac{3}{2}}$, respectively (this fact is reflected
in the units chosen for the axes). Each point on the curves in
figures 1 and 2 gives the radius and maximum possible mass of a
white dwarf when the scalar condensate extends over a region in
the stellar interior characterized by the value of the parameter
$\xi_L$ on that particular point in the curve. In figure 1, all
curves merge at point $(1,1)$ as they should since this
corresponds to the limit $\xi_L\rightarrow0$ for which no
macroscopic fraction of the star's volume is occupied by the
condensate and thus the extra dispersion force is not operative.
Figure 2, in turn, shows the other limit (i.e.
$\xi_L\rightarrow\xi_1$; point (1,1) on the top left corner) for
which the whole star is subject to the pull of the new force
(gravity being negligible). A situation that is exactly equivalent
to the Chandrasekhar case but with a blown up Newton constant. All
equilibrium configurations on the curves of figure 1 (beyond
$R\sim1.5R_{Ch}$) correspond to scalar masses in the ranges:
$8\times10^{-8}$ $eV$ $\leq m\leq2.6\times10^{-7}$ $eV$ (for
$1+\kappa=10^{2}$), $2.5\times10^{-6}$ $eV$ $\leq
m\leq6\times10^{-6}$ $eV$ (for $1+\kappa=10^{3}$),
$8\times10^{-5}$ $eV$ $\leq m\leq2\times10^{-4}$ $eV$ (for
$1+\kappa=10^{4}$), $2.5\times10^{-3}$ $eV$ $\leq
m\leq6\times10^{-3}$ $eV$ (for $1+\kappa=10^{5}$), and $0.08$ $eV$
$\leq m\leq0.2$ $eV$ (for $1+\kappa=10^{6}$). Similarly, in figure
2, the resulting scalar mass range is:
$1.4\times10^{-12}(1+\kappa)$ $eV$ $\leq
m\leq7\times10^{-12}(1+\kappa)$ $eV$. A glance at these mass
ranges\footnote{Scalar masses were obtained making $\mu_{e}=2$ and
$\rho_c=10^{8}g/cm^3$. For other choices, one should multiply the
given numerical mass values by the factor
$\left(\frac{\mu_e}{2}\right)^{\frac{4}{3}}\left(\frac{\rho_c}{10^{8}{g/cm^3}}\right)^{\frac{1}{6}}$.
} shows that the solutions in figure 1 and figure 2 exclude each
other, i.e. either we have stars of type 1 or we have stars of
type 2, but not both.

%%%%%%%%%%%%%%%%%%%%%%%%%%%%%%%%%%%%%%%%%%%%%%%%%%%%%%%%%%%%%%%%%%%%%%%%%
%%%%%%%%%  EXAMPLE OF FIGURE  %%%%%%%%%%%%%%%%%%%%%%%%%%%%%%%%%%%%%%%%%%%
\begin{figure}

\centering \psfrag{c}{$(1+\kappa)^{\frac{1}{2}} R/R_{Ch}$}
\psfrag{d}{$(1+\kappa)^{\frac{3}{2}}M/M_{Ch}$}
\includegraphics[width=5in]{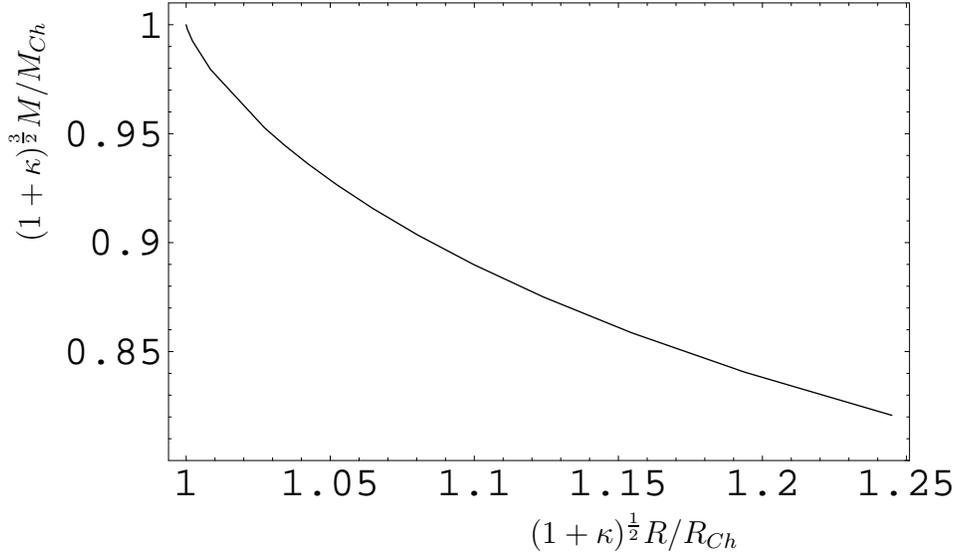}
%\hspace{1cm}
%\includegraphics[width=2.3in]{fit4.eps}
%\includegraphics[width=2.3in]{fit6.eps} \hspace{1cm}
%\includegraphics[width=2.3in]{fit8.eps}
\caption{Equilibrium configurations as a function of the parameter
$\xi_L$. As one moves along the curve from right to left, $\xi_L$
grows from $3$ to $\xi_1$. }\label{ZZZ}
\end{figure}

Next issue on our agenda is to check whether the values for
$\kappa$ and $m$ make physical sense. As to $\kappa$, we use the
specific potential in equation (12) with $\alpha\equiv m\gamma
\delta^{-2}\sim 4\pi Ga^{2}\rho_c mR^{-2}$ and use $T_c={\cal
O}(1)$ $eV$. We set the scale for the temperature $T_c$ at 1 $eV$
because with this choice the temperature of the scalars will be
below the cosmic background temperature at matter domination when
formation of structure begins. This is reasonable since the
scalars, which were in thermal equilibrium with photons in the
early stages of the history of the Universe when temperatures were
on the order of the hypothesized high energy unification scale,
have not experienced the reheating associated to the various
extinction processes (e.g., $p\bar{p}$, $\mu^{+}\mu^{-}$,
$e^{+}e^{-}$) that have otherwise risen the temperature of
photons. Little algebra and a short numerical calculation will
convince us that $g^{2}$ varies in the interval
\begin {equation}
g^{2}\sim
\left(6\times10^{-43}-2\times10^{-35}\right)\left(\frac{\rho_c}{10^8{g/cm^3}}\right)^\frac{3}{4}
\end {equation}
for the values of $\kappa$ considered. Thus, an interaction which
is extremely weak at a microscopic level is enhanced to
considerable strength by the macroscopic coherence of the scalar
medium.

 At this point one may wonder if it is realistic to think that the scalar medium can be
kept at a temperature below $1 eV$ during the formation of the
white star while the temperature of the star itself can reach
temperatures up to $10^{4}-10^{5} eV$. But, in fact, the scalars
and the stellar matter are not in thermal contact. They do not
share a common temperature. Indeed, the tiny coupling of scalars
to nucleons implies an elastic scattering cross section off non
relativistic nucleons which is at most $\sigma _{\phi N}\sim {\cal
O}(10^{-64}cm^{2})$. Clearly, $\sigma _{\phi N}^{\frac{1}{2}}$ is
so much smaller than the nucleon spacing in a nucleus, that the
nucleus is transparent and the cross section $\sigma _{\phi A}$ on
a nucleus with mass number $A$ is $\sim A$ $\sigma _{\phi N}$. As
a consequence, the largest possible collision rate of a scalar
with nuclei\footnote {With $A\simeq 20$.} in stellar matter
$\langle \sigma v n\rangle$ is so extremely small that not even in
a hundred white dwarf lifetimes a scalar would have collided once
with a nucleus. Needless to say, fixing the condensation
temperature at about $1 eV$ is by no means crucial. There is a
wide band of temperatures for which the resulting scenario is
phenomenologically sound.

Also, the effects of this scalar force would go unnoticed by local
gravity experiments. Indeed, the ratio of the vacuum potential
(equation (13); we suppose that in a terrestrial environment,
there is no appreciable amount of scalars trapped) to the
newtonian potential is
\begin {equation}
\frac{{\cal V}_{vac}}{{\cal V}_{Newt}}\simeq6.9\times
10^{-23}\left(\frac{g^{2}}{10^{-35}}\right)\left(\frac{m}{10^{-2}eV}\right)^2
\left(\frac{1}{mr}\right)^{\frac{3}{2}}{e^{-2mr}}
\end {equation}

for $r\gg m^{-1}$, and
\begin {equation}
\frac{{\cal V}_{vac}}{{\cal V}_{Newt}}\simeq1.5\times
10^{-28}\left(\frac{g^{2}}{10^{-35}}\right)\left(\frac{cm}{r}\right)^{2}
\end {equation}
for $r\ll m^{-1}$. We will see below that the domain within which
$m^{-1}$ is contained is roughly $\sim10^{-4}cm -10^{2}cm$. No
experiment searching for new forces has or will exclude in a
foreseeable future such a tiny force \cite {long}.

 A comment on $m$ comes next. The condensed phase consists of scalar
 particles with macroscopic de Broglie wavelength $\lambda_{dB}\sim
\cal{O}$$(L)$. The Compton wavelength of the particles, on the
other hand, should be much smaller than their de Broglie
wavelength for the zero-point momentum to be negligible as
compared to mass. In the low $\xi_L$ case this means:
$m\gg\lambda_{dB}^{-1}=10^{-11}-10^{-9}$ $eV$ for $(1+\kappa)$
between $10^{2}$ and $10^{6}$, respectively. The masses associated
with figure 1 show no conflict with this requirement, where even
in the most unfavorable case (for $(1+\kappa)=10^{2}$) $m$ is
almost a factor $10^{4}$ bigger than the corresponding inverse de
Broglie wavelength. In the high $\xi_L$ case one should have:
$m\gg\lambda_{dB}^{-1}=10^{-12}-10^{-10}$ $eV$ for $(1+\kappa)$
between $10^{2}$ and $10^{6}$, respectively. Obviously, the range
$1.4\times10^{-12}(1+\kappa)$ $eV$ $\leq
m\leq7\times10^{-12}(1+\kappa)$ $eV$ in figure 2, comfortably
satisfies this requirement only for sufficiently high
$(1+\kappa)$. In short, probably the mass $m$ can be reasonably
constrained to lie between $\cal O$($10^{-7}-10^{-6}$) $eV$ and
$\cal O$($0.1$) $eV$. Or, in terms of Compton wavelength:
$\sim10^{-4}cm -10^{2}cm$. Hence, the range of the 2-scalar force
for masses such as above $-$ if it were not for the condensation
phenomenon $-$ would be at most on the order of one meter. It is
only because the dispersion forces become of infinite range that
they can coherently extend over a large piece of stellar matter.

The values of $m$ entertained in the preceding paragraphs are
below or much below the temperature $T\sim1$ $eV$. Hence the
scalars are relativistic or ultrarelativistic as they should.
Furthermore, $T/m\ll Q$ as can be seen from:
\begin {equation}
Q=3.35\times10^{46}(1+\kappa)^{\frac{3}{2}}\left(\frac{10^{-2}eV}{m}\right)^{\frac{1}{2}}\left(\frac{T_c}{1
eV}\right)^{\frac{7}{2}}\left(\frac{10^{8}g/cm^{3}}{\rho_c}\right)^{\frac{3}{2}}\left(\frac{R}{R_{Ch}}\right)^{3}
\end {equation}
which follows from equation (5).

 We conclude with a summary. Two scalar exchange among nucleons produces spin independent Van der
Waals type forces. In vacuum their range is about the Compton
wavelength of the scalar particles exchanged. But when ordinary
matter is embedded in a Bose-Einstein condensate composed of the
very same mediating scalars, those forces become infinitely
ranged. The force acts then coherently over macroscopic extensions
of bulk matter. The phenomenon arises as a combination of
kinematics $-$ 3-momentum exchange of the matter system with the
scalar medium $-$ and the collective effect of condensation. The
main purpose of the present paper has been to find and study a
physical system where this phenomenon of infinite widening of the
range could be relevant.

%%%%%%%%%%%%%%%%%%%%%%%%%%%%%%%%%%%%%%%%%%%%%%%%%%%%%%%%%%%%%%%%%%%%%%%%%
%%%%%%%%%  EXAMPLE OF FIGURE  %%%%%%%%%%%%%%%%%%%%%%%%%%%%%%%%%%%%%%%%%%%
\begin{figure}

\centering \psfrag{a}{$R/R_{Ch}$} \psfrag{b}{$M/M_{Ch}$}
\includegraphics[width=5in]{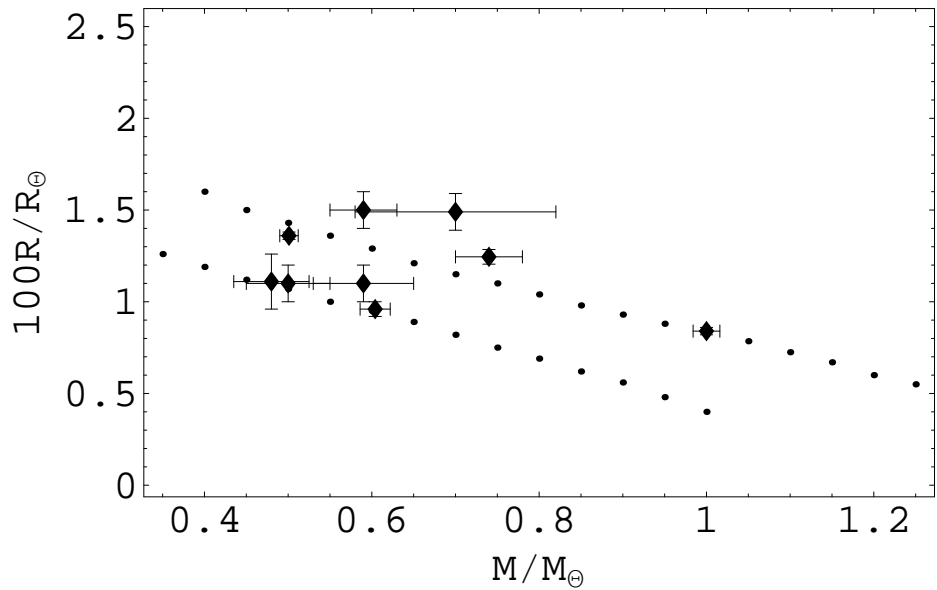}
%\hspace{1cm}
%\includegraphics[width=2.3in]{fit4.eps}
%\includegraphics[width=2.3in]{fit6.eps} \hspace{1cm}
%\includegraphics[width=2.3in]{fit8.eps}
\caption{White dwarf masses and radii. Data points correspond to
the white dwarfs Sirius B, Stein 2051, 40 Eri B and Procyon B in
binaries, and to the white dwarfs in the Common Proper-Motion
Pairs CD-38 10980, W485A, L268-92, G181-B5B and G156-64 \cite
{pro}. Theoretical calculations lie within the two dotted lines.
}\label{ZZZ}
\end{figure}

%%%%%%%%%%%%%%%%%%%%%%%%%%%%%%%%%%%%%%%%%%%%%%%%%%%%%%%%%%%%%%%%%%%%%%%%%
%%%%%%%%%  EXAMPLE OF FIGURE  %%%%%%%%%%%%%%%%%%%%%%%%%%%%%%%%%%%%%%%%%%%
\begin{figure}

\centering \psfrag{a}{$R/R_{Ch}$} \psfrag{b}{$M/M_{Ch}$}
\includegraphics[width=5in]{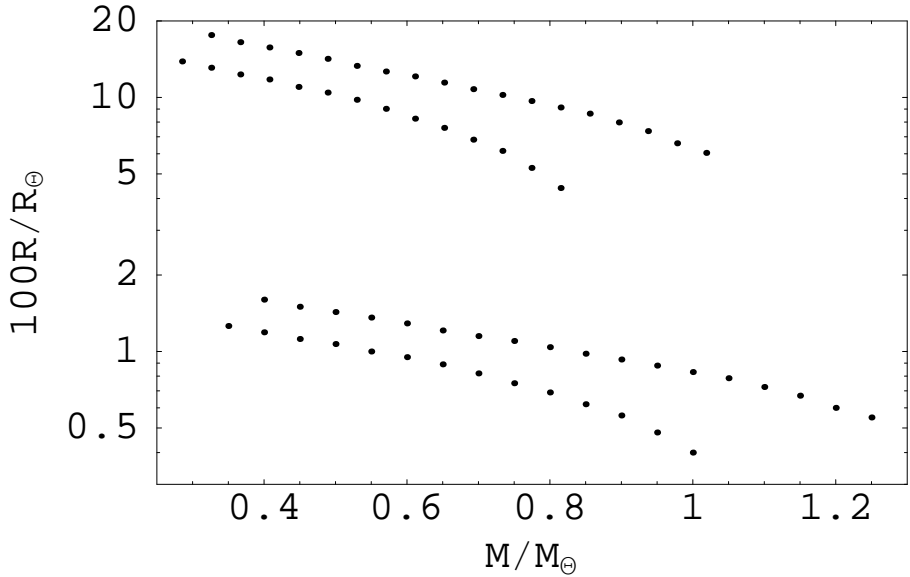}
%\hspace{1cm}
%\includegraphics[width=2.3in]{fit4.eps}
%\includegraphics[width=2.3in]{fit6.eps} \hspace{1cm}
%\includegraphics[width=2.3in]{fit8.eps}
\caption{The two dotted upper lines limit the area where type 1
stars (for the choice of parameters: $1+\kappa=10^{2}$ and
$m=1.5\times 10^{-7}$) should lie. The two dotted lower lines
correspond to the lines for actual white dwarfs in figure 3.
}\label{ZZZ}
\end{figure}

Scalars are fundamental ingredients of the SM of particle physics
and completions thereof. Quite generally, the interactions of new
matter with ordinary matter at the postulated new high energy
unification scale in those beyond the SM scenarios will render low
energy residual effects, however tiny, involving ordinary matter
at rest. In particular, the dispersion forces due to the double
exchange of scalars just mentioned will show up. These scalars, on
the other hand, may constitute a component of dark matter. If this
is the case, it might well be that chunks of scalar galactic dark
matter in the condensed phase be trapped in stellar media. As a
result, long range scalar forces would affect the equilibrium of
those stars. Restricting our analysis to white dwarfs, we have
explored the changes induced on their radii and maximal mass $-$
i.e. the Chandrasekhar limit $-$ by the presence of dispersion
forces. Although figures 1 and 2 show only this limit, in order to
relate our results to actual white dwarf data, we do the following
exercise. Figure 3 displays masses and radii for a handful of
white dwarfs obtained from visual binaries or common proper-motion
systems \cite {pro}. Also shown are two dotted lines that embrace
the different model calculations (including the pioneering
Chandrasekhar models) \cite {chan,ham}. We can get a rough idea
for where in the mass-radius plane our hypothetical stars might
lie, by scaling the band between dotted lines on figure 3 with
$\frac{M}{M_{Ch}}$ and $\frac{R}{R_{Ch}}$ on figures 1 and/or 2.
As an example let us take $1+\kappa=10^2$ and $m=1.5\times10^{-7}$
$eV$ which corresponds to a type 1 star with
$\frac{M}{M_{Ch}}=0.815$ and $\frac{R}{R_{Ch}}=11$ (see figure 1).
Figure 4 shows the resulting band as well as, for comparison, the
band for ordinary white dwarfs of figure 3. We see that, depending
on the mass $m$ of these scalars and strength $\kappa G$ of the
potential, a distinctive feature of these star configurations
would be that they should populate, relative to ordinary white
dwarfs, patently different regions on the $(M,R)$-plane. A novel
clue on dark matter might thus come from the identification of
such anomalous white dwarfs.

Work partially supported by the CICYT Research Project
FPA2002-00648, by the EU network on Supersymmetry and the Early
Universe HPRN-CT-2000-00152, and by the DURSI Research Project
2001SGR00188


\newpage

\begin{thebibliography}{99}
\bibitem{raff} See, e.g., G. Raffelt, Proc. of the 1997 European
School of High-Energy Physics, Menstrup near Naestved, Denmark,
hep-ph/9712538; J. Garcia-Bellido, astro-ph/0502139.
\bibitem{DM} K. C.Freeman, Astrophys. J. 160 (1970) 811; S. M. Faber and J. J. Gallagher, Ann. Rev. Astron.
Astrophys. 17 (1979)135; V. C. Rubin, W. K. Ford Jr., N. Thonnard,
Astrophys. J. 238 (1980) 471; D. Burstein, V. C. Rubin,  N.
Thonnard, W. K. Ford Jr., Astrophys. J. 238 (1982) 70; V. C.
Rubin, W. K. Ford Jr., N. Thonnard, D. Burstein, Astrophys. J. 261
(1982) 439; K. G. Begeman, A. H. Broeils, R. H. Sanders, Mon. Not.
R. Astr. Soc. 249 (1991) 523.
\bibitem{DM2} R. G. Carlberg et al., Astrophys. J. 462 (1996) 32;
A. Dekel, Proc. Jerusalem Winterschool 1996, Cambridge University
Press (1997).
\bibitem{oli} For a recent review see, e.g., K. A. Olive,
astro-ph/0503065.
\bibitem{wimp} N. J. C. Spooner et al., Phys. Lett. B473 (2000) 330; R. Bernabei et al., Phys. Lett. B480 (2000)
23; G. Angloher et al., Nucl. Instr. Meth. A520 (2004) 108; D. S.
Akerib et al., Phys. Rev. Lett. 93 (2004) 211301.
\bibitem{fein} G. Feinberg, J. Sucher, C. K. Au, Phys. Rep. 180
(1989) 83.
\bibitem{ferrer1} F. Ferrer, J. A. Grifols, Phys. Rev. D58 (1998) 096006.
\bibitem{ferrer2} F. Ferrer, J. A. Grifols, Phys. Rev. D63 (2000) 025020.
\bibitem{ferrer3} F. Ferrer, J. A. Grifols, JCAP 12 (2004) 012.
\bibitem{horo} C. J. Horowitz, J. Pantaleone, Phys. Lett. B319
(1993) 186.
\bibitem{haber} H. E. Haber, H. A. Weldon, Phs.Rev. Lett. 46 (1981)
1497. See also: H. E. Haber, H. A. Weldon, Phys. Rev. D25 (1982)
502; J. I. Kapusta, Phys. Rev. D24 (1981) 426; J. Bernstein, S.
Dodelson, Phys. Rev. Lett. 66 (1991) 683.
\bibitem{shapi} See: S. L. Shapiro, S. A. Teukolsky, Black Holes,
White Dwarfs, and Neutron Stars, John Wiley and Sons, Inc., New
York (1983).
\bibitem{long} J. C. Long, H. W. Chan, J. C. Price, astro-ph/9805217;
C. Hoyle, D. J. Kapner, B. R. Heckel, E. G. Adelberger, J. H.
Gundlach, U.Schmidt, H. E. Swanson, hep-ph/0405262; E. G.
Adelberger, E. Fishbach, D. E. Krause, R. D. Newman, Phys. Rev.
D68 (2003) 062002; E. G. Adelberger, B. R. Heckel, A. E. Nelson,
Ann. Rev. Nucl. Sci. 53 (2003) 77; E. Fishbach, D. E. Krause,
Phys. Rev. Lett. 83 (1999) 3593; R. S. Decca, E. Fishbach, G. L.
Klimchitskaya, D. E. Krause, D. L. Lopez, V. M. Mostepanenko,
Phys. Rev. D68 (2003) 116003.
\bibitem{pro} J. L. Provencal, Astrophys. J. 494 (1998) 759.
\bibitem{chan} S. Chandrasekhar, An Introduction
to the Study of Stellar Structure, University of Chicago Press,
Chicago, Illinois (1939).
\bibitem{ham} T. Hamada and E. E. Salpeter, Astrophys. J. 134
(1961) 683.

\end{thebibliography}
\end{document}